\documentclass[aps,prl,showpacs,superscriptaddress,twocolumn,numerical author-year,reprint]{revtex4}
\usepackage{amsfonts,amssymb,amsmath,wasysym,times}
\usepackage{color}
\usepackage{algorithm}
\usepackage{algpseudocode}
\usepackage{graphicx} 
\usepackage[usenames,dvipsnames]{xcolor}
\usepackage{pdfpages}

\usepackage{subfiles}
\begin{document}


\title{Transitivity reinforcement in the coevolving voter model}

\author{Nishant Malik}
\email{nishant.malik@dartmouth.edu}
\affiliation{Department of Mathematics, Dartmouth College, Hanover, NH 03755, USA}

\author{Feng Shi}
 \affiliation{Computation Institute, University of Chicago, Chicago, IL 60637, USA}

\author{Hsuan-Wei Lee}
\affiliation{Carolina Center for Interdisciplinary Applied Mathematics, Department of Mathematics, University of North Carolina, Chapel Hill, NC 27599, USA}

\author{Peter J. Mucha} 
\affiliation{Carolina Center for Interdisciplinary Applied Mathematics, Department of Mathematics, University of North Carolina, Chapel Hill, NC 27599, USA}

\date{\today}

\begin{abstract}

One of the fundamental structural properties of many networks is triangle closure. Whereas the influence of this transitivity on a variety of contagion dynamics has been previously explored, existing models of coevolving or adaptive network systems use rewiring rules that randomize away this important property. In contrast, we study here a modified coevolving voter model dynamics that explicitly reinforces and maintains such clustering. Employing extensive numerical simulations, we establish that the transitions and dynamical states observed in coevolving voter model networks without clustering are altered by reinforcing transitivity in the model. We then use a semi-analytical framework in terms of approximate master equations to predict the dynamical behaviors of the model for a variety of parameter settings.

\end{abstract}
\pacs{89.75.Hc, 87.23.Ge, 64.60.aq, 89.75.Fb}
\maketitle

The study of dynamics on networks has led to a number of successes identifying how the structure of the underlying network impacts the dynamics occurring on the network  \cite{centola1,barahona1,zhou1} and whether dynamics taking place on the network also promotes organizing features of the network structure itself \cite{ito1,watts1,barabasi1}. Within this larger research theme, significant attention has been given to exploring the role of network structures on the spread of contagions and opinions \cite{bogu1,thilo1,shi1,malik1}, including efforts to understand and quantify features in the spread of contagions due to different local and global structural properties \cite{may1,marceau1}. The study of opinions spreading in social networks has gained additional interest due to the rise of social media and its role in mobilizing and framing public opinion \cite{ebel1}, including elections and advertising campaigns \cite{socialmedia1}. Hence, understanding and quantifying the interplay between network structures and contagion dynamics is of very broad interest and scope \cite{centola1}.  

The processes involved in collective opinion formation, including the role of network properties in these processes, are extremely complex \cite{holme1}. We thus aim to study the properties emerging from a simple local model for interaction that incorporates only some of the essential features involved. The coevolving voter model is one of the simplest and most studied generic models for the interplay between opinion formation and the network. In this model, connected nodes with discordant opinions are resolved by one neighbor in the pair either changing its opinion or dropping the connection (in favor of a newly rewired connection to another node in the network). This model reproduces several complex features observed in collective opinion formation and has led to a variety of computational and analytical results on different aspects of the model \cite{gracia1,holme1,shi1,shi2,malik1,thilo2}. However, all previous variants of this model (including those studied by the present authors) have ignored one of the most fundamental features of networks, namely the higher propensity for a connection between two nodes that are both already connected to a third node, closing the triangle between them \cite{watts1}. Specifically, the rewiring rules in these models (and in a wide variety of other adaptive networks models) ignore clustering, pushing the network structure further towards independently-distributed edges (up to the coupling with node states). The probability of closing a triangle along a potential edge in a connected triple converges over time in these models to the same probability for that edge in the absence of those other connections. That is, only trivial levels of local clustering are observed. As such, the applicability of such models for describing real systems is highly questionable. 

In this Letter, we introduce a new variation of the coevolving voter model that explicitly reinforces transitivity to generalize to networks with more realistic local clustering. We modify the rewiring step to preferentially rewire to neighbors of neighbors, mimicking the common social phenomena that friends of friends are more likely to be friends. We provide the details of the model and introduce different end states that can result from the two-opinion model with reinforced transitivity. We then analyze the structural properties of the evolving networks and transitions. Finally, we use approximate master equations (AME) to predict model behavior in different parameter regimes. Additional details about the analytical derivations and numerical experiments are provided in the accompanying supplementary material (SM).

\begin{figure}
 \centering
 \includegraphics [width=0.95\columnwidth]{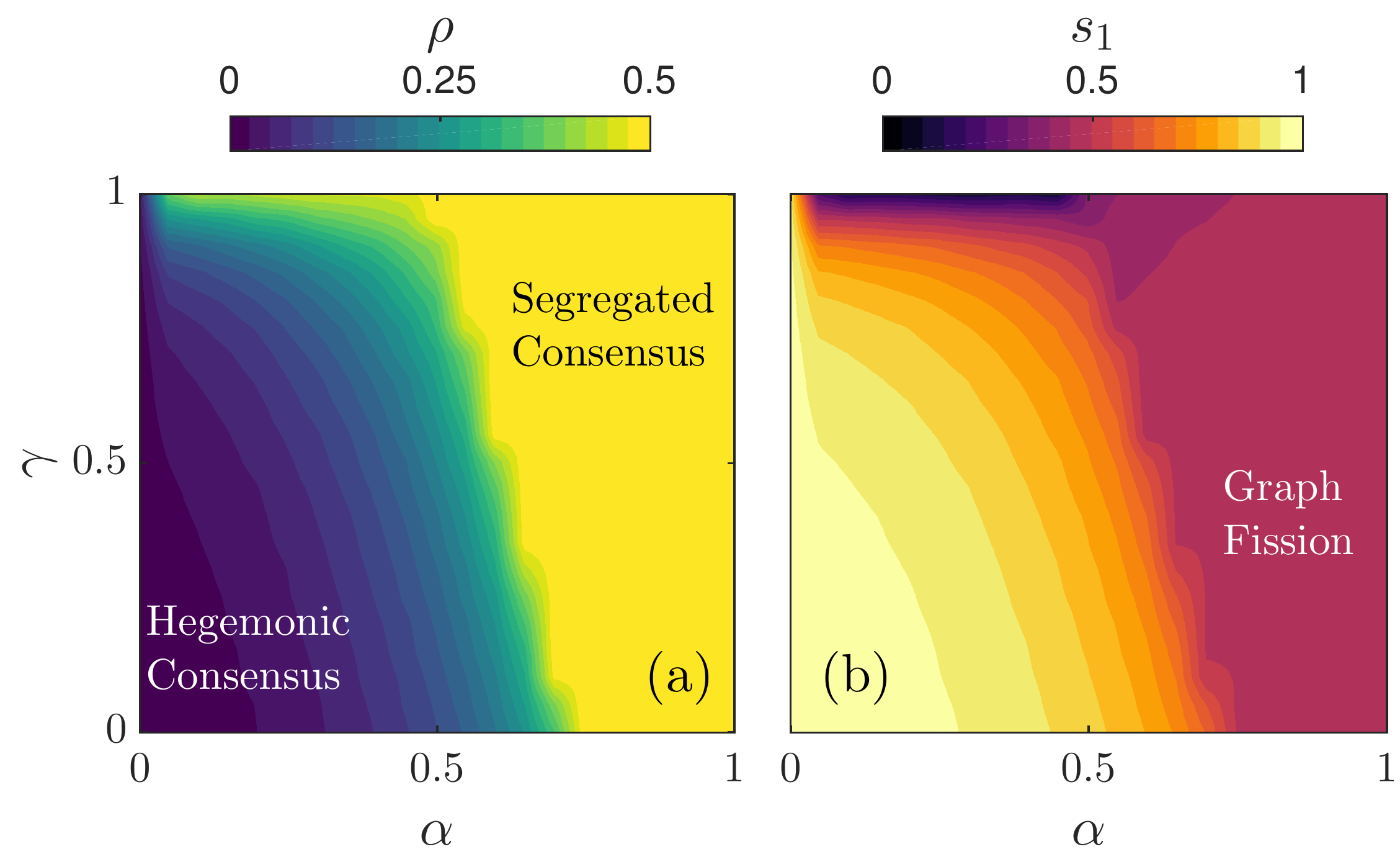}
 \caption{(Color online) (a) The fraction of nodes holding the minority opinion in the consensus state, $\rho$. (b) The fraction of nodes in the largest connected component at consensus, $s_1$. Both (a) and (b) indicate that the critical rewiring probability $\alpha_c(\gamma)$ decreases with increasing triangle closure probability $\gamma$. Below $\alpha_c(\gamma)$, the segregated consensus gives way to an ever more hegemonic consensus with decreasing $\alpha$.}
  \label{fig:phplt1}
  \end{figure}

Consider a graph $G$ with $N$ nodes and $l$ edges (links) with each node holding one of two opinions (0 and 1). We call an edge \emph{discordant} if it connects nodes with different opinions and let $l_{01}$ be the fraction of edges that are discordant. Similarly, let $l_{00}$ and $l_{11}$ be the fractions of the two types of \emph{harmonious} edges (connecting nodes with the same opinion).  At each step of the model process, a discordant edge is chosen. With probability $1-\alpha$, a node at one end will adopt the other's opinion; with probability $\alpha$, one of the nodes breaks this link and rewires to another node. The essential reinforcement of transitivity occurs in this rewiring step: with probability $\gamma$, the new neighbor is selected from the set of nodes two steps away---that is, neighbors of neighbors---if this set is non-empty; otherwise (with probability $1-\gamma$ or, in the probability $\gamma$ case if the neighbors-of-neighbors set is empty), a node is selected uniformly at random from the network. The total number of edges at time $t$ is conserved, with $l =l_{01} (t)+l_{00}(t)+l_{11}(t)=1$. In the simulations presented here, we use a network with $N=100,000$ nodes, to minimize finite size effects, with average degree $\langle k \rangle=2l/N=4$, initialized as an Erd\H{o}s-R\'{e}nyi  random graph with the two opinions distributed uniformly over the nodes with each opinion selected with probability $1/2$. 

The final state of the model reached at $t=t_f$ is a \emph{consensus state} with $l_{01}(t_f) =0$, i.e.\ there are no discordant edges remaining and no further evolution of the system takes place. 
We loosely classify consensus states into two broad categories: \emph{hegemonic} and \emph{segregated}, based on the fraction of nodes holding the minority opinion at consensus, $\rho$. The hegemonic consensus is characterized by small $\rho$; in contrast, the segregated consensus is characterized by minimal change in the populations of the two opinions, $\rho\approx 0.5$ [see Fig.~\ref{fig:phplt1}(a)], with opinions distributed in separate connected components that are each in internal consensus. Similarly, in Fig.~\ref{fig:phplt1}(b) we see that the fraction of nodes in the largest connected component at consensus, $s_1$, is approximately 0.5 in the segregated consensus and increases with decreasing $\alpha$ as the consensus becomes more and more hegemonic. 

Generalizing from the ``rewire-to-random" model in \cite{shi1}, corresponding to the $\gamma=0$ case here, and noting the relatively small changes with increasing $\gamma$ in most of Fig.~\ref{fig:phplt1}, we expect the consensus state to be qualitatively consistent with \cite{shi1} for small triangle-closing probability $\gamma$, with a critical value for the rewiring probability, $\alpha_c(\gamma)$, above which only a segregated consensus state exists. Below $\alpha_c(\gamma)$, the consensus becomes more and more hegemonic for decreasing $\alpha$. The argument $\gamma$ in $\alpha_c(\gamma)$ signifies the dependence on the tendency to close triangles in rewiring. As observed in Fig.~\ref{fig:phplt1}, this critical value $\alpha_c(\gamma)$ appears to decrease consistently with increasing $\gamma$ before sharply changing as $\gamma$ gets closer to $1$. This transitivity reinforcement is thus important in altering the dynamics of the coevolving voter model, yet appears to preserve many of the qualitative features of the consensus states, at least for $\gamma$ not too close to $1$. 
   
To better understand the role of triangle closure on the dynamics, in Fig.~\ref{fig:evolC1}(a) we plot the evolution of transitivity in simulations with $\alpha=1$ (no opinion switching). As our results show, even after initializing with an Erd\H{o}s-R\'{e}nyi  random graph, we see that transitivity reinforcement causes transitivity to increase over time in these simulations, except in the $\gamma=0$ (no reinforcement) case, with larger $\gamma$ driving larger transitivity. The transitivity in the consensus states, $\mathcal{C}(t_f)$, is highlighted in the Figure by circles. Using a simple mean field argument that assumes convergence to statistically stationary levels of transitivity (see SM section S1), we estimate $\mathcal{C}(t_f)= 3\gamma/(3\langle k \rangle-2)$. Even though the clustering is still increasing with time in Fig.~\ref{fig:evolC1}(a), we observe in Fig.~\ref{fig:evolC1}(b) that this theoretical estimate matches well with the clustering coefficients at consensus in the simulations.  

\begin{figure}
 \includegraphics [width=\columnwidth]{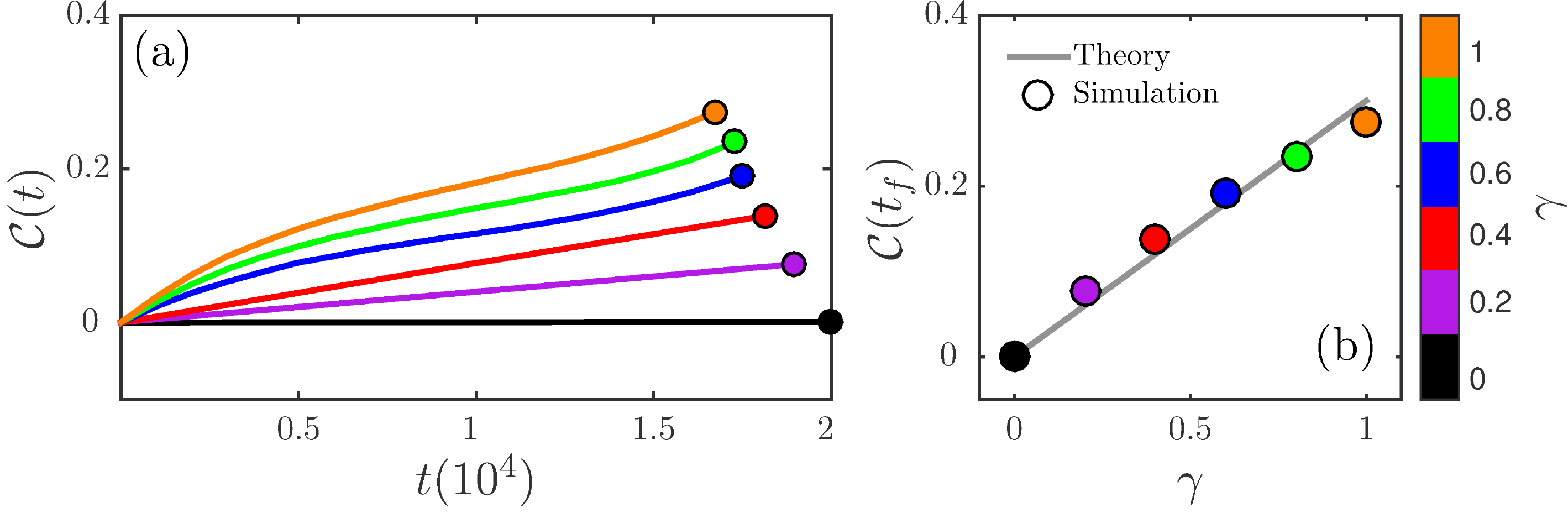}
 \caption{\small { (Color online) (a) Time evolution of clustering coefficient in $\alpha=1$ (no opinion switching) simulations for different $\gamma$. Circles highlight the clustering coefficient in the consensus state, i.e., $\mathcal{C} (t_f)$.  (b) The value of clustering in the consensus state, comparing simulations (circles) and the theoretical estimate.  See also Fig.~S1 for comparison at other $\alpha$ values, demonstrating good agreement with the theory except for small $\alpha$ with large $\gamma$.}}
  \label{fig:evolC1}\end{figure}

The interplay of opinion changes (without rewiring) and the rewiring steps alters the degree distribution of the network. In Fig.~\ref{fig:deg1}, we illustrate the variations induced in the degree distribution of the consensus state at different $(\alpha,\gamma)$ values. At $\alpha=0$, there is no rewiring and the degree distribution at consensus is the same as the initial Poisson degree distribution (grey bands in Fig.~\ref{fig:deg1}). For $\alpha=0.2$ and $\alpha=0.4$, the consensus degree distribution deviates more and more from the initial distribution as $\gamma$ is increased. Whereas each random rewiring step can only maintain or decrease the number of discordant edges, an opinion switching step net increase the total amount of disagreement, slowing down the convergence to consensus \cite{holme1,shi1,shi2,malik1}. For smaller $\alpha$, it typically takes more steps to reach consensus, giving greater opportunity for increased $\gamma$ (closing a greater number of triangles) to cause deviations in the degree distribution. For $\alpha>\alpha_c(\gamma)$, we observe only a minor departure from the initial degree distribution, even for higher values of $\gamma$, as the rewiring step dominates and the graph quickly disintegrates into connected components that are each in internal consensus. Indeed, the number of steps for segregated consensus is $O(N\log N)$ (for given average degree) \cite{shi1},  yielding fewer rewiring steps overall and limiting the total change in the degree distribution.

\begin{figure}
 \centering
 \includegraphics [width=0.95\columnwidth]{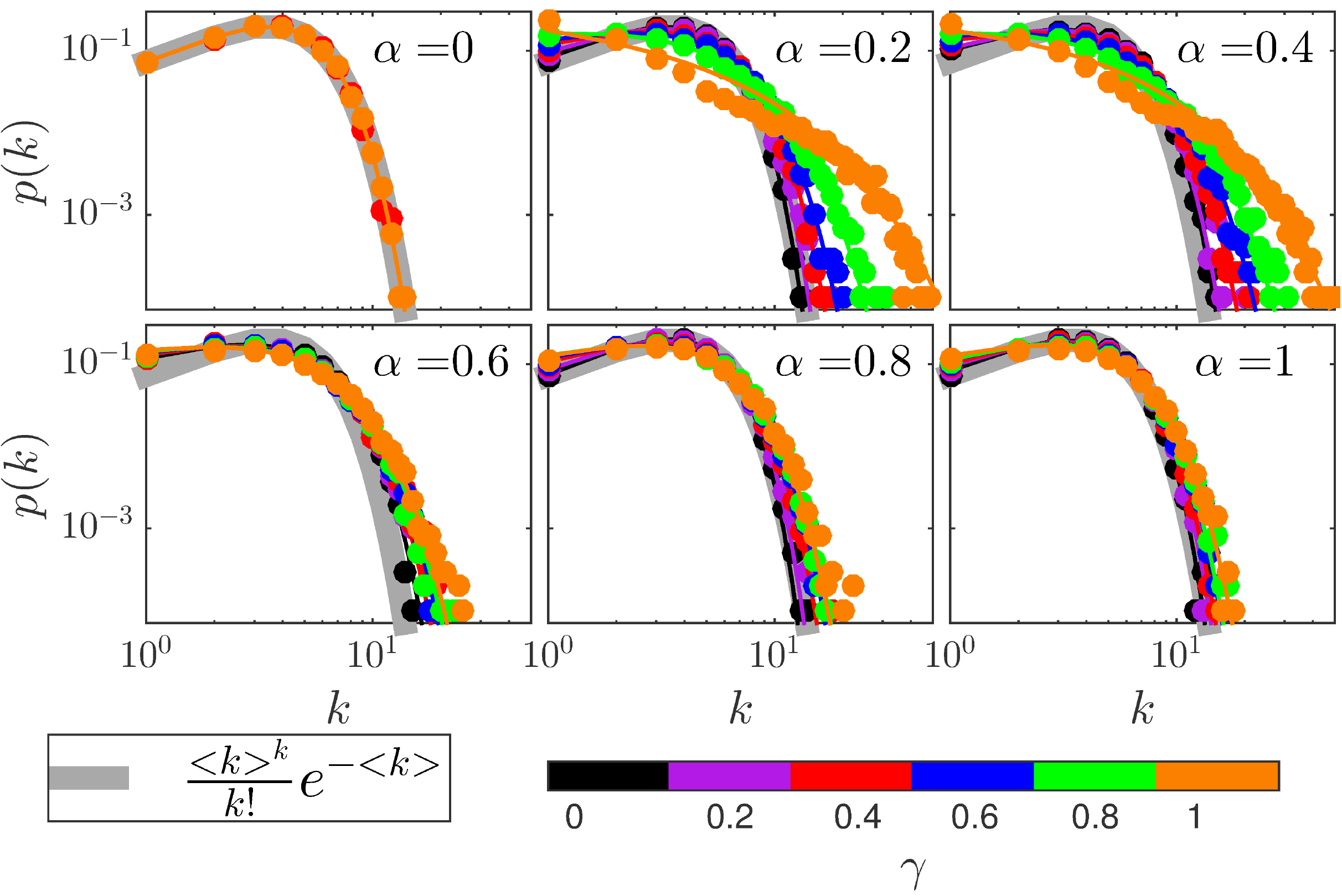}
 \caption{\small (Color online) {Degree distribution in the consensus state for different $(\alpha,\gamma)$ parameters. Simulations start from Erd\H{o}s-R\'{e}nyi random networks with $\langle k \rangle=4$, with Poisson degree distribution indicated by thick grey bands.}}
    \label{fig:deg1}
\end{figure}

To further quantify the influence of $\gamma$ on consensus degree distribution, we have identified the following fit to the data plotted in Fig.~\ref{fig:deg1}:
\begin{eqnarray}
p(k) = \begin{cases} \displaystyle  \frac{\langle k \rangle^k }{k!}e^{-\langle k \rangle}, & \mbox{if } \alpha = 0\,,  \\[1em] 
  \displaystyle \frac{b_1}{1.25\langle k \rangle} \left(\frac{k}{1.25\langle k \rangle}\right)^{b_1-1}e^{ -(\frac{k}{1.25\langle k \rangle})^{b_1}}, & \mbox{if }  \alpha \neq 0\,,  \end{cases}
\label{eq:eqx2}
\end{eqnarray}
where the $\alpha \neq 0$ cases are fit by Weibull distributions with shape parameter $b_1$ and scale parameter fixed constant equal to $1.25\langle k \rangle$ (see also Figs.~S2 and S3). The Weibull distribution is used here to capture the additional observed variance compared to the initial Poisson degree distribution.  The values of the shape parameter $b_1$ are plotted in Fig.~S3. In particular, we observe lower values of $b_1$ for $\alpha=0.2$ and $0.4$ as compared to $\alpha=0.8$ and $1.0$.

In discussing Fig.~\ref{fig:phplt1}, we noted the transition between the hegemonic and segregated consensus states in terms of the critical parameter $\alpha_c(\gamma)$, extending its identification in \cite{shi1} to the $\gamma>0$ transitivity reinforcing dynamics considered here. Further generalizing \cite{shi1}, we observe that the fraction of discordant edges at time $t$, $l_{01}(t)$, for $\alpha < \alpha_c(\gamma)$ obeys an approximate relationship describing a family of quasi-stationary states that behave as attracting sets for the dynamics, with
${ l_{01}(t)=c_1(1-n_1(t))n_1(t)+c_2 }$, 
where $n_1(t)$ is the fraction of nodes holding opinion $1$, and $c_1$ and $c_2$ are constant-in-time values dependent on $(\alpha,\gamma)$. Solving the quadratic equation for the $l_{01}=0$ consensus state yields $n_{1_{\pm}}=\frac{1}{2}(1 \pm \sqrt{ 1+ 4c_2/c_1})$,  where $n_{1_+}$ (respectively, $n_{1_-}$) represents the state when $n_1$ is the majority (minority) opinion. That is, $\rho=\frac{1}{2}(1 - \sqrt{ 1+ 4c_2/c_1})$.  In Fig.~\ref{fig:trs1}(a-b), we plot the arches approximated by these parabolae. As $\alpha$ and  $\gamma$ are increased, these arches disappear for $\alpha > \alpha_c(\gamma)$.  In  Fig.~\ref{fig:trs1}(b),  we observe that as $\gamma$ is increased the arches become squeezed, decreasing the area enclosed under the arches.  

  \begin{figure*}
 \centering
 \includegraphics [width=2.0\columnwidth]{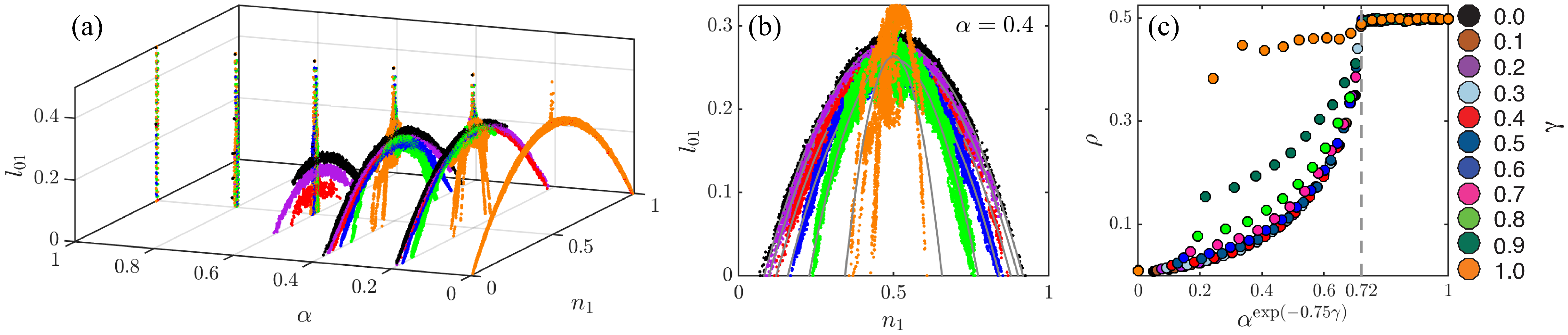}
 \caption{\small { (Color online) (a) Simulation dynamics in the space of variables $\l_{01}$ and $n_1$. The trajectories of $\l_{01}$ and $n_1$ rapidly relax to these arches. As $\alpha$ and $\gamma$ are varied, the shape of the arches change, disappearing for higher $\alpha>\alpha_c(\gamma)$.  (b) $\l_{01}$ vs $n_1$ for $\alpha=0.4$. Observe the squeezing of arches as $\gamma$ is increased, breaking up for $\gamma=1.0$.  (c) The minority opinion population $\rho$ for different $\gamma$ and $\alpha$.  The abscissa has been transformed to  $\alpha^{\exp (-0.75 \gamma)}$ to provide a common location for the critical point near $0.72$ after rescaling, while collapsing most of the data for $\gamma<0.8$ onto a single curve.}}
    \label{fig:trs1}
    \end{figure*}

Estimates for $c_1$ and $c_2$ from the simulation data [see Fig.~\ref{fig:trs1}(a-b)] are plotted in Fig.~S4. We also observe (in Fig.~S5) that the ratio of these coefficients appearing in the formula for $\rho$ above approximately follows ${c_2/c_1  \approx -\frac{1}{2}\alpha^{2.1 \exp (-0.75 \gamma)}}$. Using this observation of the fitted arch parameters to identify the $\gamma$ dependence of $\alpha_c(\gamma)$, in Fig.~\ref{fig:trs1}(c) we plot $\rho$ vs.\ $\alpha^{\exp (-0.75 \gamma)}$ for different $\alpha$ and $\gamma$. As evident from the figure, $\alpha_c(\gamma) \approx (0.72)^{\exp(0.75\gamma)}$ accurately quantifies the shift in $\alpha_c$ with $\gamma$. Moreover, from Fig.~\ref{fig:trs1}(c) we observe that this rescaling of the $\rho$ vs.\ $\alpha$ relationship below the critical value falls onto nearly the same curve for $\gamma<0.8$.

Further details about the quasi-stationary states may be approximated through reduced-order model equations. Mean field and pair approximation methods are popular tools for describing binary state dynamics on networks, but have been found inadequate in many complex models \cite{jglesson}. A more powerful approach is in terms of Approximate Master Equations (AME), with coupled differential equations describing the evolution of binary states of nodes and their neighbors \cite{jglesson}. We have generalized the AME equations of \cite{shi1} for $\gamma>0$ transitivity reinforcement, as presented in section S4 of the SM. In Fig.~\ref{fig:simanly1}, we compare the quasi-stationary states predicted by AME with those observed in simulations. Importantly, we note that the discrepancy between the AME and simulation arches already present at $\gamma=0$ (in agreement with \cite{shi1}) increases slightly as $\gamma$ is increased but nevertheless captures the main changes as long as $\gamma$ is not too large.

\begin{figure}
 \centering
 \includegraphics [width=0.8\columnwidth]{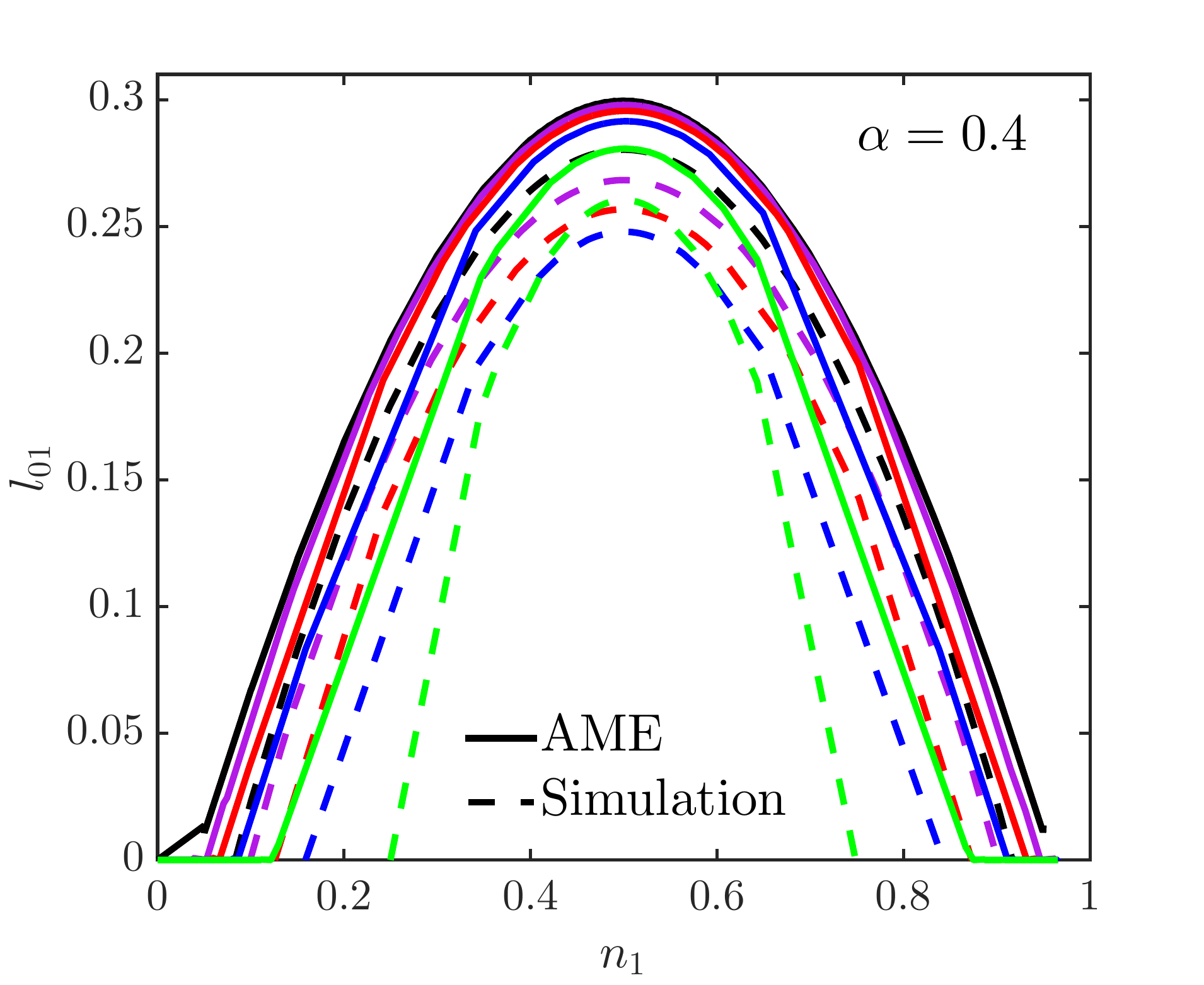}
 \caption{\small { (Color online)  Comparison of Approximate Master Equation (AME) solutions with simulations at $\alpha=0.4$. Different colors represents different values of $\gamma$ by the same color scheme as in Figs.~2--4. Simulation results presented here correspond to the arches fitted to raw simulation data as shown in Fig.~\ref{fig:trs1}(b). }}
    \label{fig:simanly1}
    \end{figure}

In summary, we observe that multiple features of our new transitivity-reinforcing model show continuous transitions in the consensus states, in qualitative but not precise quantitative agreement with the model without transitivity reinforcement studied in \cite{shi1} (corresponding to $\gamma=0$ here). Importantly, we have found that the critical value for these transitions depends on the extent of transitivity reinforcement in the model. We thus conclude that reinforcement of clustering alters the internal details of the coevolving voter model in terms of reaching consensus and shifting the critical transitions. Therefore, one should be careful in interpreting applicability of results based on models without clustering. We also demonstrate that the method of approximate master equations can be used in this setting to predict the impact of transitivity reinforcement in shifting the macroscopic properties of the dynamics and the resulting consensus.

\acknowledgements

N. Malik, Hsuan-Wei Lee, and P. J. Mucha acknowledge support from Award Number R21GM099493 from the National Institute of General Medical Sciences and Award Number R01HD075712 from the Eunice Kennedy Shriver National Institute of Child Health \& Human Development. Feng Shi acknowledges support from the John Templeton Foundation to the Metaknowledge Network. The content is solely the responsibility of the authors and does not necessarily represent the official views of the funding agencies. 
\bibliographystyle{unsrt}


\newpage

\setcounter{equation}{0}
\setcounter{figure}{0}
\setcounter{table}{0}

\renewcommand{\thefigure}{S\arabic{figure}}
\renewcommand{\thetable}{S\arabic{table}}
\renewcommand{\thesection}{S\arabic{section}}
\renewcommand{\theequation}{S\arabic{equation}}

\title{Supporting Material: Transitivity reinforcement in the coevolving voter model}

%
%
%


\maketitle

\onecolumngrid

\section{S1. Mean field estimate for the evolution of clustering in the model}

Let $T$ be the number of triangles and $\tau$ be the number of connected triplets of nodes (triads) in the network at a given time $t$. Then the global clustering coefficient will be $\mathcal{C}(t)=3T/ \tau$. Further, let $\tau_{j}$ be the number of triads centered at node $j$. We note that $\tau_{j} = \dbinom{k_j}{2}$, where $k_j$ is the degree of node $j$.   If during the rewiring step a link is removed from node $j$ and rewired to node $m$, the number of triads centered at $j$ reduces by $\dbinom{k_j}{2}-\dbinom{k_j-1}{2} = k_j-1$, while the number of triads centered at $m$ increases by $\dbinom{k_m+1}{2}-\dbinom{k_m}{2} = k_m$. Then the total change in the number of triads in a single rewiring step is $\Delta \tau = k_m-k_j+1.$  Assuming (without justification) that the degrees of the nodes before losing and gaining the rewired link are independent and identically distributed (\emph{iid}), then on average the change in the number of triads per rewiring step is $\langle \Delta \tau \rangle  =1$.  

The rewiring rate at given time $t$ is proportional to the probability of rewiring, $\alpha$, and we scale time so that the expected instantaneous rate of change of $\tau$ will be (on average, abusing notation for simplification) $\dot{\tau}= \alpha \langle \Delta \tau \rangle = \alpha$. We remark that we have scaled time here per consideration of any discordant edge. An alternative is to scale time so that every discordant edge is considered on average once per unit time, introducing multiplicative factors of the number of discordant edges above and in what follows in such way that they cancel and do not affect the steady state. We thus ignore these factors in what follows. 
 
The rewiring step also changes the number of triangles  $T$  in the network.  Let $T_{ij}$ be the number of triangles which include the edge $i$--$j$. If this edge is removed during the rewiring then $T_{ij}$ triangles will be eliminated.  There are two types of triads involved with edge $i$--$j$: the $k_i-1$ ones centered at node $i$ and the $k_j-1$ others centered at node $j$. That is, the total number of triads involved with edge $i$--$j$ is $k_i+k_j-2$. We note that this count of these triads includes each of the $T_{ij}$ triangles twice. We additionally note that each of the $T_{ij}$ triangles associated with the $i$--$j$ edge is by definition associated with two other edges. Then, using the fact that the clustering coefficient $\mathcal{C}$ represents the fraction of triads that are involved in triangles, and assuming independence and uniformity throughout, we obtain $T_{ij}=\mathcal{C}{(k_i+k_j-2)/2}$ as our estimate for the number of triangles that will be eliminated in removing the $i$--$j$ edge. Again assuming that node degree is \emph{iid}, on average the number of triangles removed per rewiring event will be $\mathcal{C} (\langle k  \rangle-1)$.

Reinforcing transitivity is the counter mechanism that rewiring to a neighbor's  neighbor occurs with probability $\gamma$. Continuing to assume uniformity and independence throughout the present argument (as just one for example, ignoring 4-cycles that might exist including both the old and new edges), then each such step increases the number of triangles by 1. That is, triangles are added by this mechanism at rate $\alpha\gamma$. 

Combining these mechanism, we write the expected net instantaneous rate of change of $T$ as  
\begin{equation}
\dot{T} = - \alpha \mathcal{C} (\langle k  \rangle-1) +  \alpha \gamma\,.
\end{equation}
From $\mathcal{C} =3T/\tau$, the statistically-steady level of clustering ($\dot{\mathcal{C}}=0$) is obtained when $\dot{T}  \tau - T \dot{\tau} =0$, giving
$\mathcal{C} = 3T/\tau =3\dot{T}/\dot{\tau}$. After substituting in the rates above, this becomes $\mathcal{C}=3\alpha[\gamma-\mathcal{C}(\langle k\rangle-1)]/\alpha$. Solving for $\mathcal{C}$ we then obtain  
\begin{equation}
\mathcal{C}=\frac{3\gamma}{3\langle k  \rangle -2}.
\label{eq:CMF}
\end{equation}

\begin{figure}
\centering 
 \includegraphics [width=0.95\columnwidth]{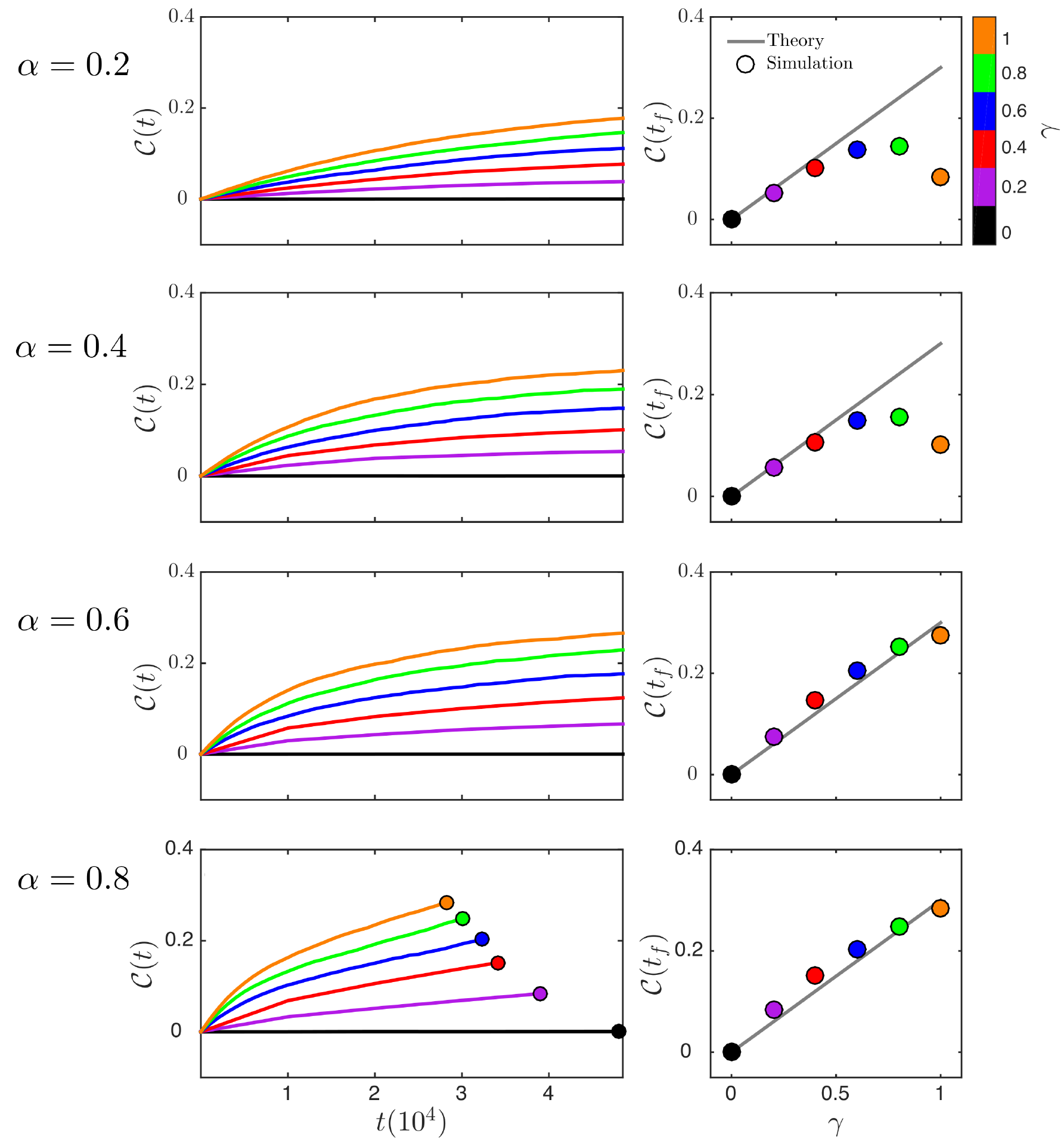}
 \caption{\small Left panels: Temporal evolution of clustering coefficient for different $\alpha,\gamma$ parameters. Right panels: The value of the clustering coefficient in the consensus state, comparing simulations (circles) and the theoretical estimate in Eq.~\ref{eq:CMF}}
  \label{fig:clustfig}
  \end{figure}

In Fig.~\ref{fig:clustfig} we plot the clustering coefficients over time and at consensus for different $(\alpha,\gamma)$ values, similar to the $\alpha=1$ data presented in Fig.~2. Note the slightly longer time scale in the left panels in Fig.~\ref{fig:clustfig} compared to Fig.~2, and that consensus is not reached on the plotted time scale for smaller values of $\alpha$. The right panels plot the final value $\mathcal{C}(t_f)$, demonstrating good agreement with Eq.~\ref{eq:CMF} except for at the larger values of $\gamma$ at smaller $\alpha$.

\begin{figure}
\centering 
 \includegraphics [width=0.95\columnwidth]{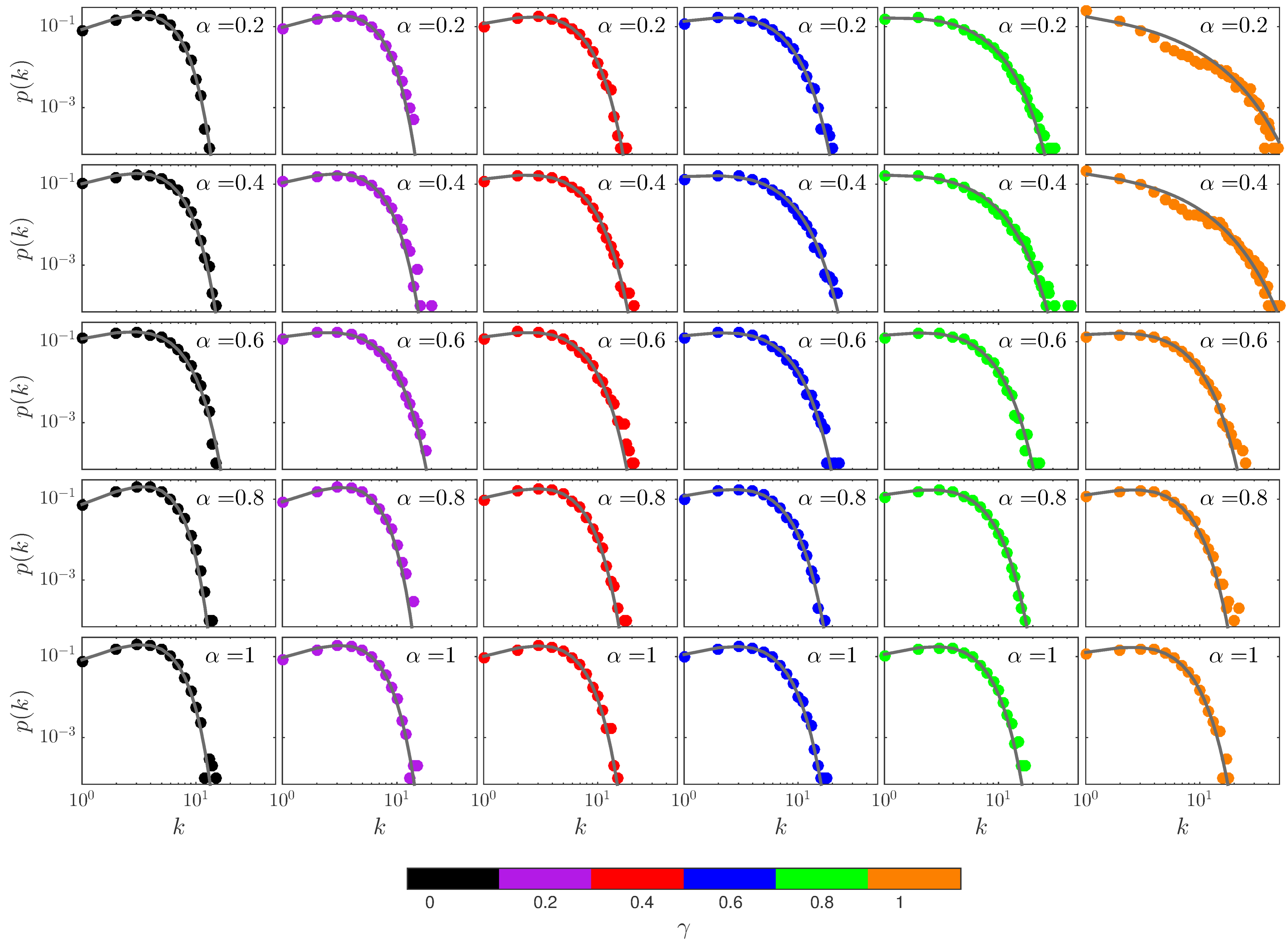}
 \caption{\small Fits to the degree distributions in the final consensus state for different values of $\alpha$ and $\gamma$. See also Eq.~1,  Fig.~3 , and Fig.~\ref{fig:b1fig}.}
  \label{fig:fit1fig}\end{figure}
\begin{figure}
\centering 
 \includegraphics [width=0.75\columnwidth]{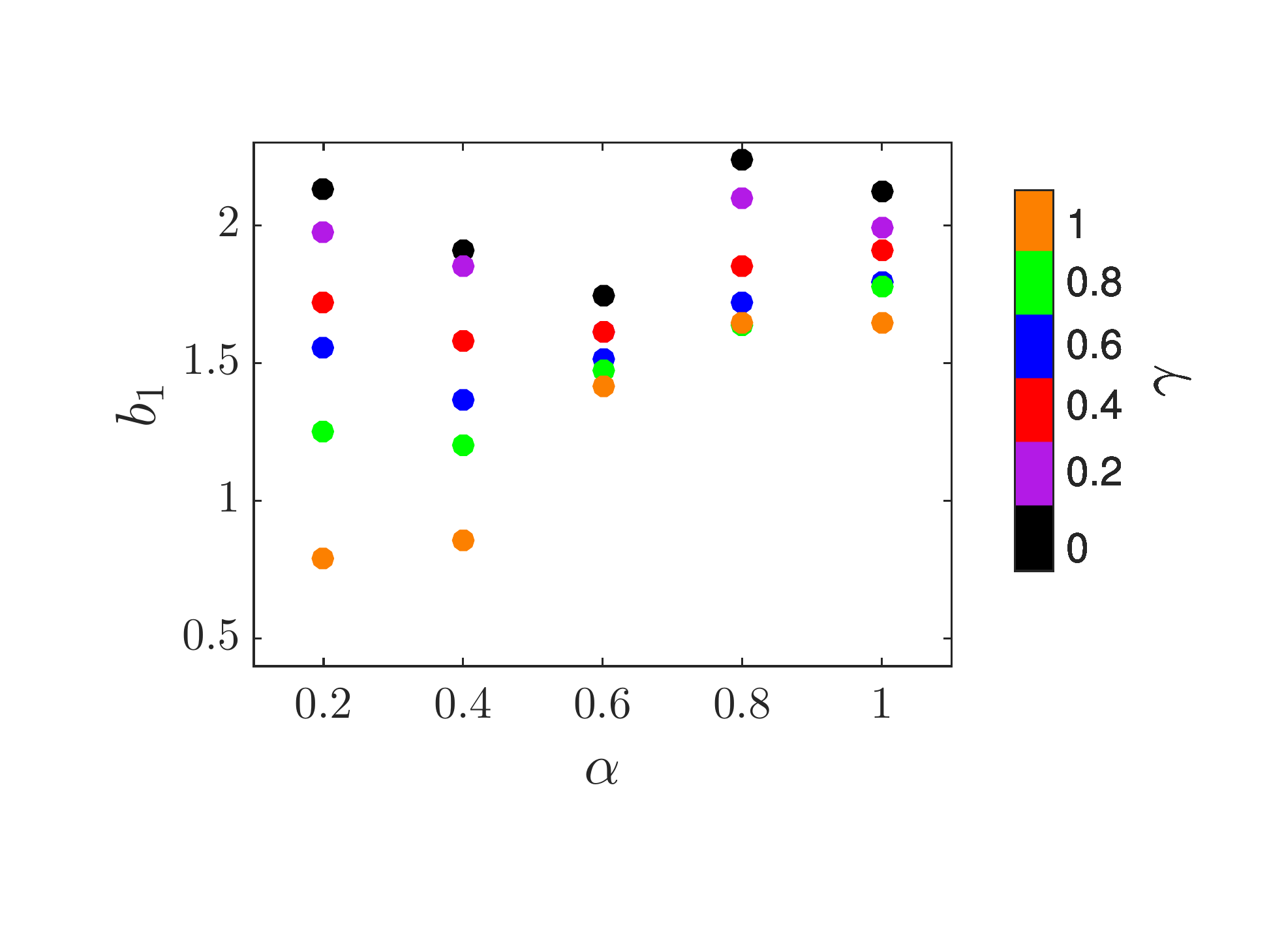}
 \vspace*{-0.7in}
 \caption{\small Values of the Weibull shape parameter $b_1$ obtained by fitting degree distributions at consensus (see Eq.~1 and Fig.~\ref{fig:fit1fig}).}
  \label{fig:b1fig}\end{figure}

\section{S2. Degree Distributions}

As rewiring is introduced into the model (that is, $\alpha>0$), the structure of the network evolves.  We observe that the degree distribution for $\alpha>0$ can be fitted by Weibull distributions (see Eq.~1 and Fig.~\ref{fig:fit1fig}). In Fig.~\ref{fig:b1fig} we plot the shape parameter $b_1$ used to fit Eq.~1. We observe bigger dispersion in the values of $b_1$ for small $\alpha$'s (see $\alpha=0.2$ and $0.4$ in the Figure). Larger values of the $b_1$ shape parameter give a larger spread of the degree distribution. In other words, neither $\alpha$ nor $\gamma$ change the fundamental character of the distribution; rather, their combination merely stretches or contracts the spread of the degree distribution.  It appears that there are two regimes in the values of  $b_1$, coinciding with $\alpha$ above and below the critical values $\alpha_c(\gamma)$. These two regimes also correspond to two different time scales involved in the evolution of the system: it takes a larger number of steps to reach consensus for $\alpha$ below the critical value. We also note that at these values of the shape parameter, the mean of the Weibull distribution is very close to proportional to its scale parameter, fixed constant equal to $1.25\langle k\rangle$ in our fits here, corresponding well to the fact that $\langle k\rangle$ remains constant.

\section{S3. Characterizing the Quasi-Stationary States}

\begin{figure}
\centering 
\includegraphics [width=0.85\columnwidth]{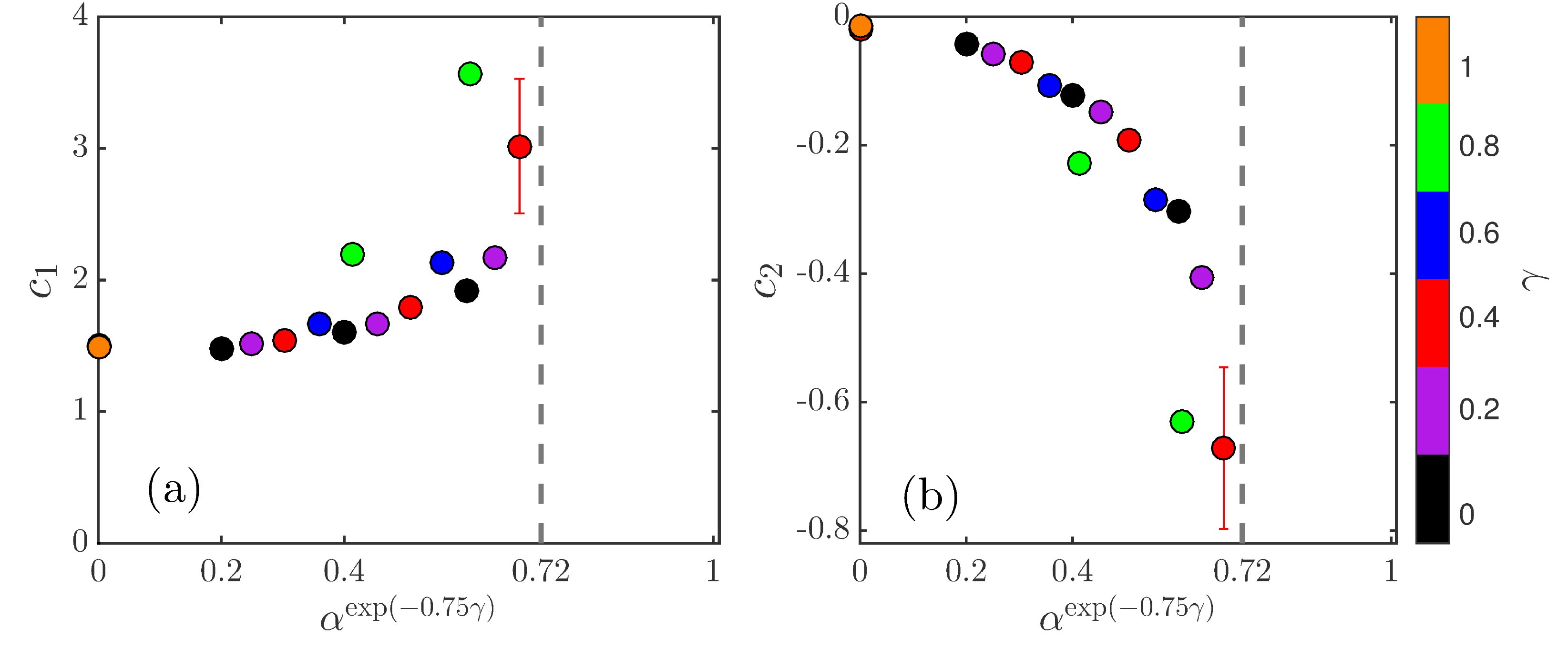}
 \caption{\small The fitted constants describing the quasi-stationary states (see Eq.~\ref{eq:eql01S}). Points where the mean squared error of the polynomial fit is greater than $0.001$ have been removed. Error bars indicate the 3$\sigma$-standard error in the estimate of $c_1$ and $c_2$.} 
 \label{fig:cfig}
  \end{figure}

\begin{figure}
\centering 
 \includegraphics [width=0.85\columnwidth]{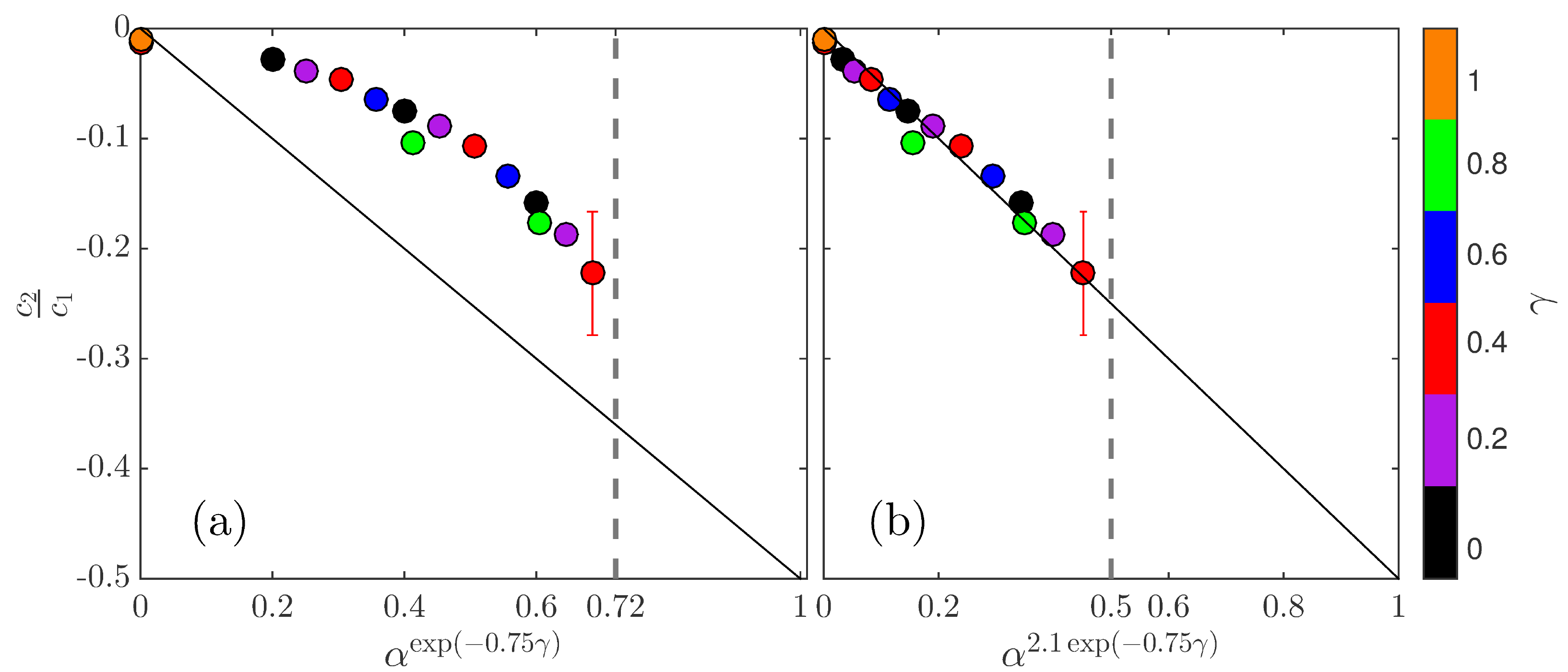}
 \caption{\small (a) The ratio $c_2/c_1$ plotted against the scalings used in Fig. 4(c), where $c_1$ and $c_2$ are the parameter estimates for the quadratic polynomial fitted to the arches in Fig. 4(a) and (b) (see Eq.~\ref{eq:eql01S}). Points where the mean squared error of the polynomial fit is greater than $0.001$ have been removed. Error bars indicate the 3$\sigma$-standard error in the ratio $c_2/c_1$. (b) The ratio replotted to demonstrate approximately linear dependence with $\alpha^{2.1 \exp (-0.75\gamma)}$ for $\alpha<\alpha_c(\gamma)$.}
  \label{fig:c1c2N1fig}\end{figure}

The quasi-stationary states appear to be attracting in the observed dynamics, in qualitative agreement with the observations in \cite{shi1,shi2} (which correspond to the $\gamma=0$ dynamics considered here). In these quasi-stationary states, the fraction of edges that are discordant, $l_{01}$, is well approximated by 
\begin{equation} 
l_{01}(t)=c_1(1-n_1(t))n_1(t)+c_2 \label{eq:eql01S} 
\end{equation} 
where $n_1$ is the fraction of nodes holding opinion 1, and $c_1$ and $c_2$ are constants over time (depending on the parameters $\alpha$ and $\gamma$). The values of $c_1$ and $c_2$ can be estimated directly from the simulation data, such as that in Fig.~4(a), as plotted here in Fig.~\ref{fig:cfig}. In so doing, we observe a fitting form for combining the dependence on $\alpha$ and $\gamma$ through the single value $\alpha^{\exp(-0.75\gamma)}$. Moreover, we observe a simple linear relationship approximating the ratio of the two constants: $c_2/c_1 \approx -\frac{1}{2} \alpha^{2.1 \exp (-0.75\gamma)}$ for $\alpha<\alpha_c(\gamma)$, as demonstrated in Fig.~\ref{fig:c1c2N1fig}.

\begin{figure}
\centering 
 \includegraphics [width=0.85\columnwidth]{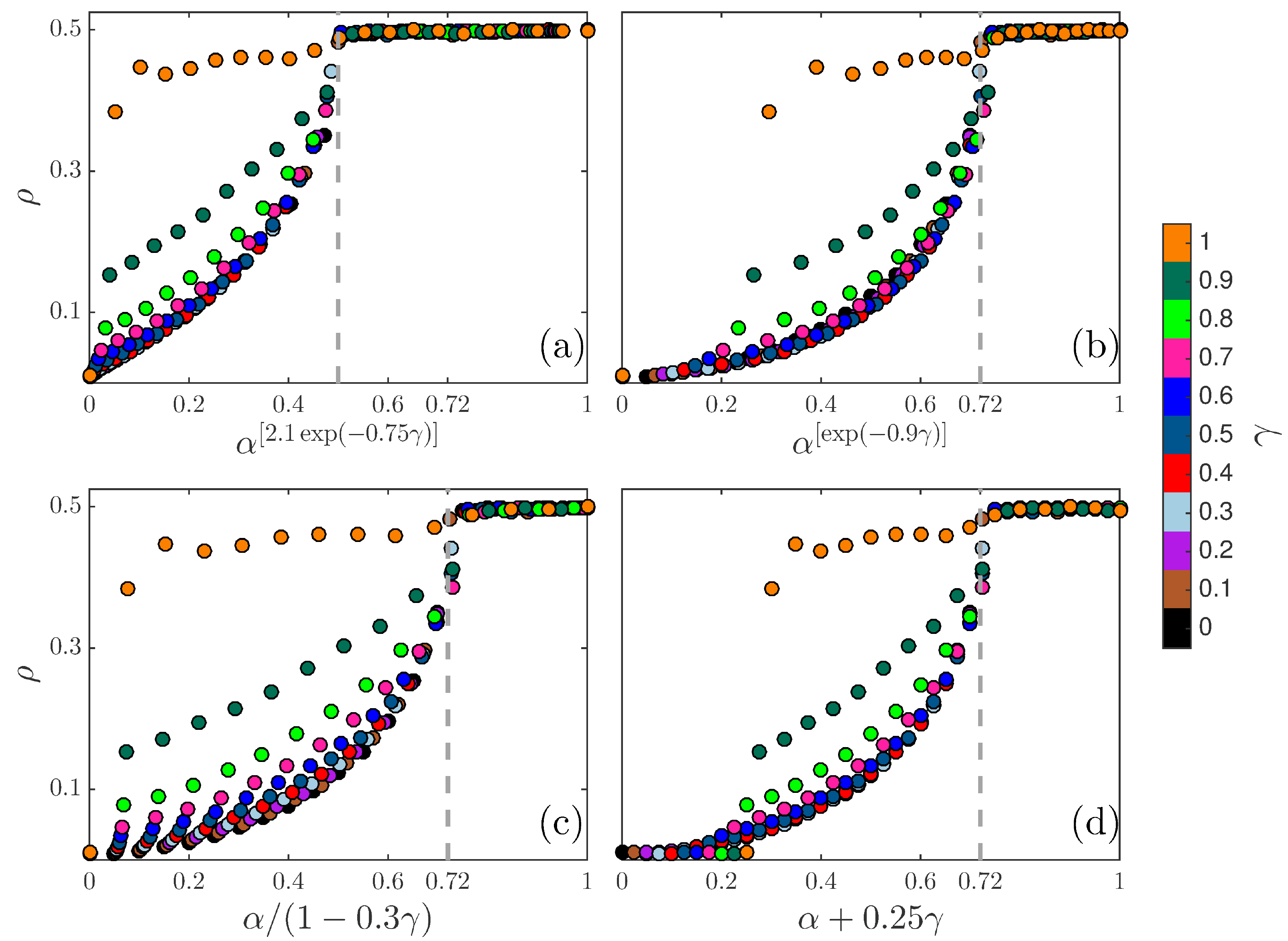}
 \caption{\small Alternative scalings for the parameter $\alpha$ with $\gamma$, representing various levels of agreement with the placement of the critical value (dashed line) and with the collapse of the data below the critical value.}
  \label{fig:c1c21fig}
  \end{figure}

Carrying forward from these observations for the fitted values of $c_1$ and $c_2$, we plot the fraction holding the minority opinion at consensus, $\rho$, versus the rescaled quantity $\alpha^{\exp(-0.75\gamma)}$ in Fig.~4c. In particular, the Figure demonstrates the good agreement with $\alpha_c(\gamma)\approx (0.72)^{\exp(0.75\gamma)}$. For comparison and completeness, in Fig.~\ref{fig:c1c21fig} we consider other possible scalings of $\alpha$ with $\gamma$, demonstrating different levels of agreement with the critical value and with the overall collapse of the curve for $\alpha<\alpha_c(\gamma)$.

\section{S4. Approximate Master Equations (AME)}

In the evolving voter model reinforcing transitivity, we introduce effects due to a node rewiring to its neighbor's neighbor. Specifically, after selecting a discordant edge, the probability of rewiring (versus opinion switching) is $\alpha$, and then within the decision to rewire the probability of a node rewiring to its neighbor's neighbor is given by the parameter $\gamma$. That is, among all steps of the model, the probability (that is, the rate) or rewiring to a neighbors' neighbor is $\alpha\gamma$, while the probability to rewire to a node at random is $\alpha(1-\gamma)$.

For the purposes of this Section, let $n_{0}$ be the fraction of nodes with opinion $0$, $n_{1}$ the fraction of nodes with opinion $1$, $l_{ab}$ the number of $a$--$b$ oriented links, and $\tau_{abc}$ the number of $a$-$b$-$c$ oriented triples having opinions $a$, $b$ and $c$, with $a,b,c \in {\{0,1\}}$. Note that in this notation, $l_{01} = l_{10}$, and $l_{00}$ counts every unoriented $0$-$0$ link twice.  Let $S_{k,m}(t)$  be the fraction of nodes with opinion $0$ that have $k$ neighbors, $m$ of which hold opinion $1$, at time $t$. Similarly, let $I_{k,m}(t)$ be the fraction of nodes with opinion $1$ that have $k$ neighbors, $m$ of which hold opinion $1$.  We follow \cite{jglesson,shi1} to develop differential equations describing the evolution of the quantities $S_{k,m}(t)$  and  $I_{k,m}(t)$.

We note that $S_{k,m}(t)$ and $I_{k,m}(t)$ conserve the number of nodes, with
\begin{equation}   \sum_{k,m} S_{k,m}(t)+ \sum_{k,m} I_{k,m}(t)=1,  \end{equation} 
and conserve the number of edges, with
\begin{equation}  
\sum_{k,m} k S_{k,m}(t) + \sum_{k,m} k I_{k,m}(t) =\langle k \rangle.
\end{equation}

If fraction $\epsilon$ of the nodes are initially (at $t=0$) made to hold opinion $1$ uniformly at random, then the initial conditions for $S_{k,m}$ and $I_{k,m}$ are given by $$S_{k,m}(0) = (1-\epsilon) p_{k}(0) \binom{k}{m}\epsilon^{m}(1-\epsilon)^{k-m},$$
and $$I_{k,m}(0) = \epsilon p_{k}(0) \binom{k}{m}\epsilon^{m}(1-\epsilon)^{k-m},$$ where $p_k(0)$ is the initial degree distribution. In order to match our simulations, $p_k(0)$ is a Poisson distribution with mean $\langle k\rangle=4$, and we set $\epsilon=0.5$.

To write the differential equation governing the evolution of $S_{k,m}$, we will require an estimate for the probability of center node in the $S_{k,m}$ count having a neighbor's neighbor (distance-2 neighbor) with opinion $0$. We denote this probability by $P(nn0 \mid S_{k,m})$ and estimate it as
\begin{align*}
 P(nn0 \mid S_{k,m})  &=  \frac{m-1}{k-1}\cdot \frac{l_{10}}{\frac{1}{2}\l_{11}+l_{10}}+\frac{k-m}{k-1}\cdot \frac{\frac{1}{2}l_{00}}{\frac{1}{2}l_{00}+l_{01}}. \\
 \end{align*}
Similarly, in our equations we similarly need this probability for center node in the $S_{k,m+1}$ count, given as
 \begin{align*}
  P(nn0 \mid S_{k,m+1})  &=  \frac{m}{k-1}\cdot \frac{l_{10}}{\frac{1}{2}\l_{11}+l_{10}}+\frac{k-m-1}{k-1}\cdot \frac{\frac{1}{2}l_{00}}{\frac{1}{2}l_{00}+l_{01}}. \\
\end{align*}
Using these quantities, our AME ODE governing the time evolution of the $S_{k,m}$ compartment is
\begin{align}
\frac{d}{dt} S_{k,m} & = \alpha \gamma  \big\{-\big[1+P(nn0 \mid S_{k,m})\big]mS_{k,m}+P(nn0 \mid S_{k,m+1})(m+1)S_{k,m+1}+(m+1)S_{k+1,m+1}\big\}  \nonumber \\
                              & + \alpha (1-\gamma) \big\{ -(2-u)mS_{k,m}+(1-u)(m+1)S_{k,m+1}+(m+1)S_{k+1,m+1}\big \}  \nonumber \\
                              & + \alpha \gamma \Big\{-\big[\big(\frac{m}{k}\cdot \frac{l_{10}}{\frac{1}{2}\l_{11}+l_{10}}\big) \cdot \frac{l_{01}}{N_{0}} + \big(\frac{k-m}{k}\cdot \frac{\frac{1}{2}l_{00}}{\frac{1}{2}l_{00}+l_{01}}\big) \cdot \frac{l_{01}}{N_{0}} \big] \cdot S_{k,m}  \nonumber \\ 
                              & -  \big[\big(\frac{m}{k}\cdot \frac{\frac{1}{2}\l_{11}}{\frac{1}{2}\l_{11}+l_{10}}\big) \cdot \frac{l_{10}}{N_{1}} + \big(\frac{k-m}{k}\cdot \frac{l_{01}}{\frac{1}{2}l_{00}+l_{01}}\big) \cdot \frac{l_{10}}{N_{1}} \big] \cdot S_{k,m} \nonumber \\
                             & + \big[\big(\frac{m-1}{k-1}\cdot \frac{\frac{1}{2}\l_{11}}{\frac{1}{2}\l_{11}+l_{10}}\big) \cdot \frac{l_{10}}{N_{1}} + \big(\frac{k-m}{k-1}\cdot \frac{l_{01}}{\frac{1}{2}l_{00}+l_{01}}\big) \cdot \frac{l_{10}}{N_{1}} \big] \cdot S_{k-1,m-1} \nonumber \\ 
                              & + [\big(\frac{m}{k-1}\cdot \frac{l_{10}}{\frac{1}{2}\l_{11}+l_{10}}\big) \cdot \frac{l_{01}}{N_{0}} + \big(\frac{k-m-1}{k-1}\cdot \frac{\frac{1}{2}l_{00}}{\frac{1}{2}l_{00}+l_{01}}\big) \cdot \frac{l_{01}}{N_{0}}\big] \cdot S_{k-1,m}\Big\} \nonumber \\
                              & + \alpha (1-\gamma) \frac{l_{01}}{N} \big\{-2S_{k,m}+S_{k-1,m-1}+S_{k-1,m}\big\} \nonumber \\
                              & + (1-\alpha)\big\{-mS_{k,m}+(k-m)I_{k,m}\big\}  \nonumber \\
                              & + (1-\alpha)\big\{-\beta^{s}(k-m)S_{k,m}+\beta^{s}(k-m+1)S_{k,m-1}-\gamma^{s}mS_{k,m}+\gamma^{s}(m+1)S_{k,m+1}\big\}
\label{eq:eq_for_skm}
\end{align}
where 
\begin{align*}
\beta^{s} &= \frac{\sum_{k,m}mS_{k,m}}{\sum_{k,m}S_{k,m}} = \frac{\tau_{001}}{l_{00}} \\
\gamma^{s} &= \frac{\sum_{k,m}(k-m)^{2}I_{k,m}}{\sum_{k,m}(k-m)I_{k,m}} = \frac{\tau_{010}}{l_{01}} + 1\,. 
\end{align*}
That is, $\beta^{s}$ is the  number of $1$ neighbors of a $0$-$0$ edge and $\gamma^{s}$ gives the  number of $0$ neighbors of the $1$ at the end of a $0$-$1$ edge and the $+1$ counts the $0$ on the conditioning edge. 

\begin{figure}[!ht]
  \centering
    \includegraphics[width=0.8\textwidth]{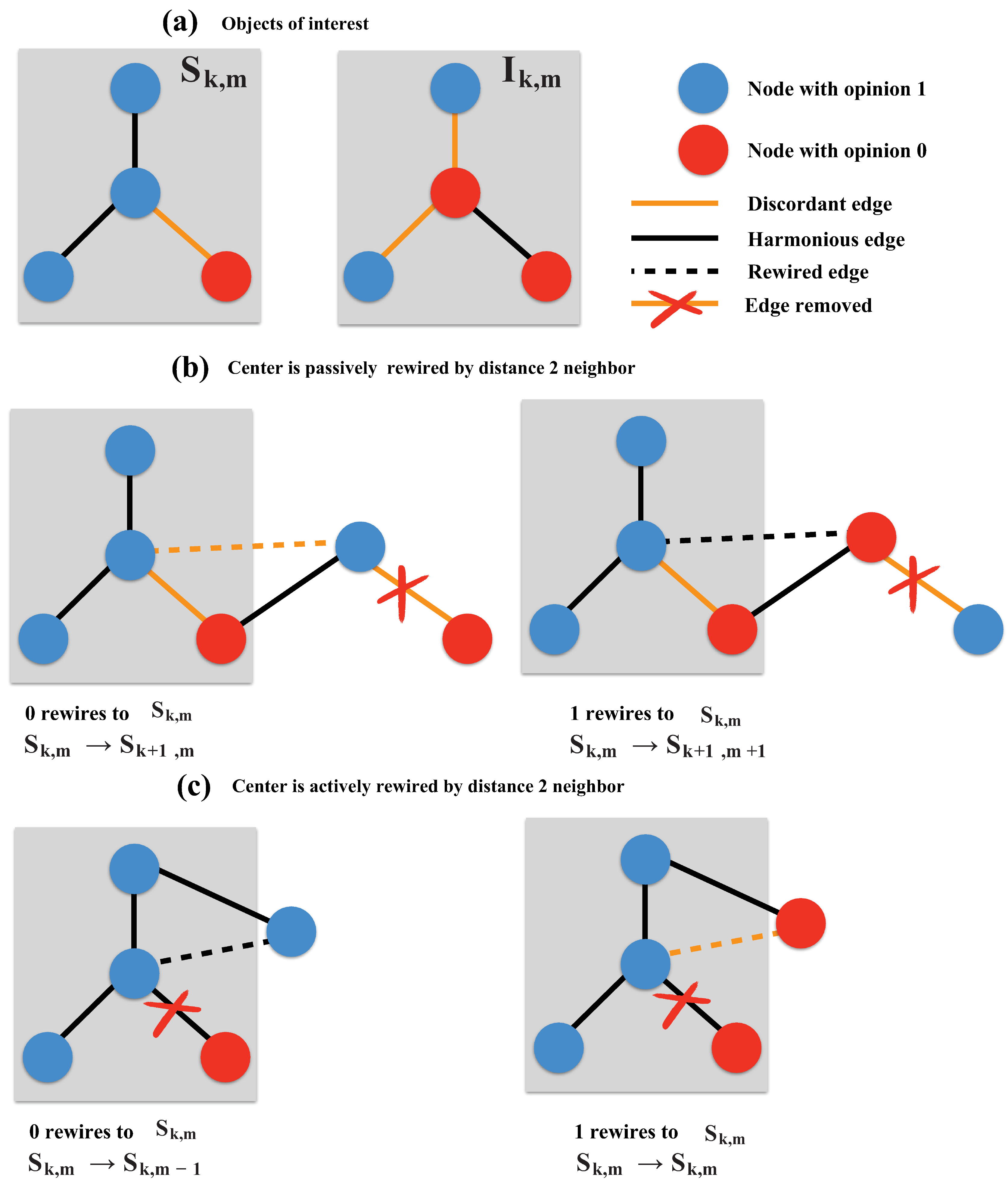}
    \caption{ Illustration enumerating some of the steps involved in construction of Eq.~\ref{eq:eq_for_skm}. (a) The two objects of interest are $S_{k,m}$ and $I_{k,m}$ (see Eq.~\ref{eq:eq_for_ikm} for $I_{k,m}$). Panels (b) and (c) shows some of the sample steps that lead to changes in the $S_{k,m}$  population. Panel (b) considers the case when a node at distance 2 rewires to the center (passive rewiring), while (c) considers the case where the center actively rewires to a node at distance 2.}
\label{fig:diarewire}
\end{figure}

The first line of the right hand side of Eq. (\ref{eq:eq_for_skm})  accounts for the case when the center actively rewires to a distance-2 neighbor. The second line accounts for the case when the center actively rewires to other nodes in the network. The third to sixth lines account for the cases when the center is passively rewired by its distance-2 neighbors. The seventh line counts the events when the center is passively rewired by other nodes in the network. Finally, the last two lines represent the voter step---i.e., no rewiring happens and the nodes simply update their opinions. Fig.~\ref{fig:diarewire} illustrates some of these rewiring steps. 

We similarly obtain the following differential equation  governing the evolution of the $I_{k,m}$:
\begin{align}
\frac{d}{dt} I_{k,m} & = \alpha \gamma \{-[1+P(nn1|I_{k,m})](k-m)I_{k,m}+P(nn1|I_{k,m-1})(k-m+1)I_{k,m-1}+(k-m+1)I_{k+1,m}\}  \nonumber \\
                              & + \alpha (1-\gamma) \big\{ -(1+u)(k-m)I_{k,m}+u(k-m+1)I_{k,m-1}+(k-m+1)I_{k+1,m}\big \}  \nonumber \\
                              & + \alpha \gamma \Big\{-\big[\big(\frac{m}{k}\cdot \frac{l_{10}}{\frac{1}{2}\l_{11}+l_{10}}\big) \cdot \frac{l_{01}}{N_{0}} + \big(\frac{k-m}{k}\cdot \frac{\frac{1}{2}l_{00}}{\frac{1}{2}l_{00}+l_{01}}\big) \cdot \frac{l_{01}}{N_{0}} \big] \cdot I_{k,m}  \nonumber \\ 
                              & -  \big[\big(\frac{m}{k}\cdot \frac{\frac{1}{2}\l_{11}}{\frac{1}{2}\l_{11}+l_{10}}\big) \cdot \frac{l_{10}}{N_{1}} + \big(\frac{k-m}{k}\cdot \frac{l_{01}}{\frac{1}{2}l_{00}+l_{01}}\big) \cdot \frac{l_{10}}{N_{1}} \big] \cdot I_{k,m}   \nonumber \\ 
                              & + \big[\big(\frac{m-1}{k-1}\cdot \frac{\frac{1}{2}\l_{11}}{\frac{1}{2}\l_{11}+l_{10}}\big) \cdot \frac{l_{10}}{N_{1}} + \big(\frac{k-m}{k-1}\cdot \frac{l_{01}}{\frac{1}{2}l_{00}+l_{01}}\big) \cdot \frac{l_{10}}{N_{1}} \big] \cdot I_{k-1,m-1}  \nonumber \\ 
                              & + \big[\big(\frac{m}{k-1}\cdot \frac{l_{10}}{\frac{1}{2}\l_{11}+l_{10}}\big) \cdot \frac{l_{01}}{N_{0}} + \big(\frac{k-m-1}{k-1}\cdot \frac{\frac{1}{2}l_{00}}{\frac{1}{2}l_{00}+l_{01}}\big) \cdot \frac{l_{01}}{N_{0}}\big] \cdot I_{k-1,m}\Big\}  \nonumber \\
                              & + \alpha (1-\gamma) \frac{l_{01}}{N} \big\{-2I_{k,m}+I_{k-1,m-1}+I_{k-1,m}\big\}  \nonumber \\
                              & + (1-\alpha)\{-(k-m)I_{k,m}+mS_{k,m}\}  \nonumber \\
                              & + (1-\alpha)\{-\beta^{i}(k-m)I_{k,m}+\beta^{i}(k-m+1)I_{k,m-1}-\gamma^{i}mI_{k,m}+\gamma^{i}(m+1)I_{k,m+1}\}
\label{eq:eq_for_ikm}
\end{align}
where 
\begin{align*}
P(nn1 \mid I_{k,m})  &= \frac{m}{k-1}\cdot \frac{\frac{1}{2}\l_{11}}{\frac{1}{2}\l_{11}+l_{10}}+\frac{k-m-1}{k-1}\cdot \frac{l_{01}}{\frac{1}{2}l_{00}+l_{01}} \\
P(nn1 \mid I_{k,m-1})  &= \frac{m-1}{k-1}\cdot \frac{\frac{1}{2}\l_{11}}{\frac{1}{2}\l_{11}+l_{10}}+\frac{k-m}{k-1}\cdot \frac{l_{01}}{\frac{1}{2}l_{00}+l_{01}}\,.
\end{align*}

\begin{figure*}
  \centering
    \includegraphics[width=0.8\textwidth]{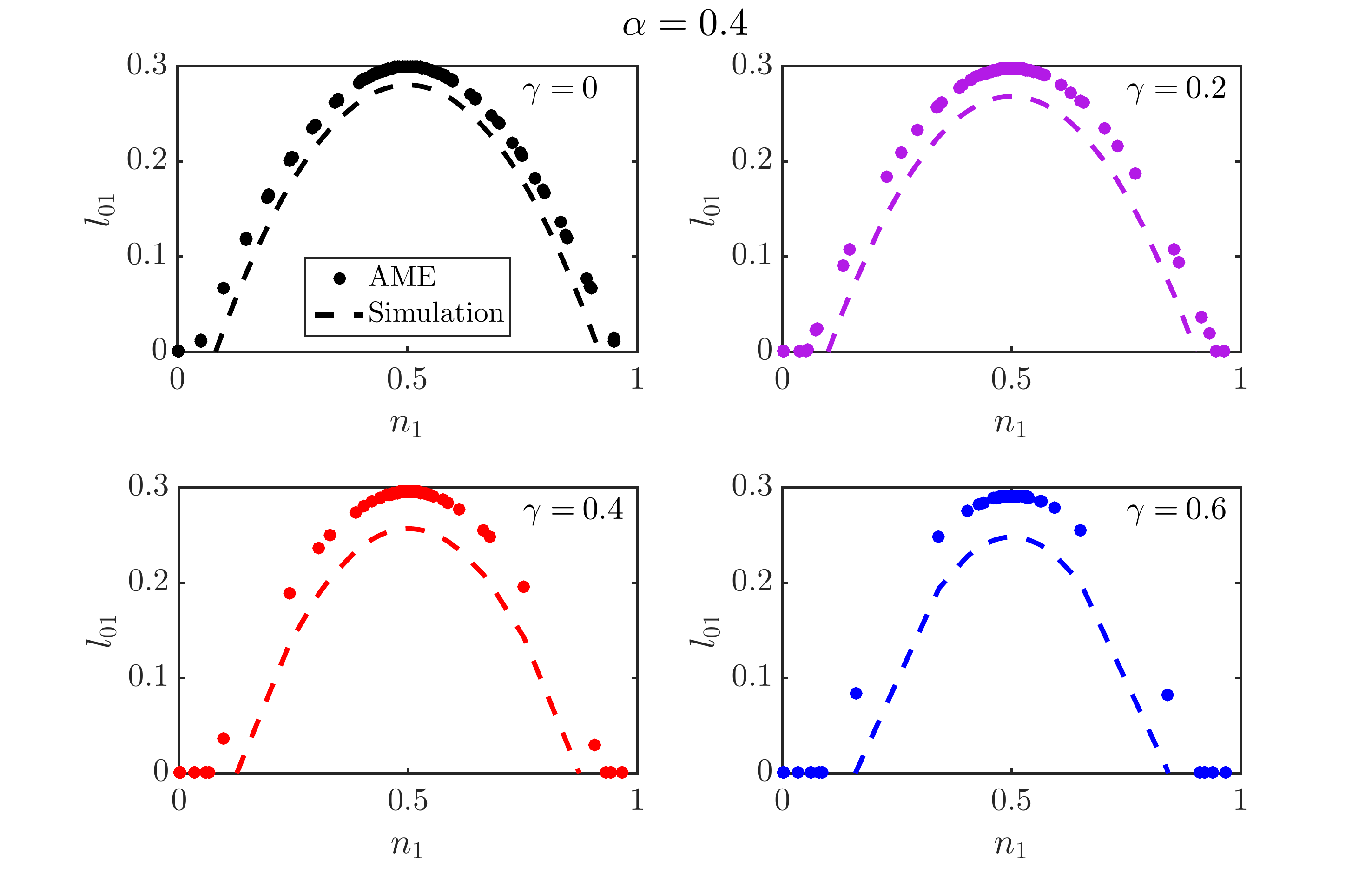}
    \caption{Comparison between Approximate Master Equation (AME)  and simulations for $\alpha=0.4$. Solutions for AME were sampled at $t=500, 1000, 2000$. }
\label{fig:alp04}
\end{figure*}

There are thus $2(k_{max}+1)^2$ equations governing the evolution of $S_{k,m}(t)$ and $I_{k,m}(t)$, where $k_\mathrm{max}$ is the maximum degree allowed in the system. That is, all populations above this maximum degree are fixed at zero; we here set $k_\mathrm{max}=20$. We numerically solve these equations using MATLAB\textsuperscript{\textregistered}'s ode45 solver was used up to times after which the observed evolution is significantly slower. From the quasi-steady populations obtained by these numerical solutions, we plot the fraction of discordant edges versus the fraction of nodes with opinion $1$ in Fig.~\ref{fig:alp04} for $\alpha=0.4$ and different $\gamma$ values, comparing with simulation results. We note in particular that the discrepancy between the AME predictions and the observed simulation behavior increases slightly as $\gamma$ increases.

\bibliographystyle{unsrt}


\end{document}